\documentclass[12pt, a4paper]{article}
\usepackage[top=1in, bottom=1in, left=1.25in, right=1.25in]{geometry}
\usepackage{booktabs}
\usepackage{multirow}
\usepackage{setspace}
\usepackage{natbib}
\bibpunct[:]{(}{)}{,}{a}{}{,}
\usepackage{authblk}
\usepackage{setspace}
\usepackage{amsmath}
\usepackage{amsthm}
\usepackage{amsfonts}
\usepackage{ascmac}
\usepackage{amssymb}
\usepackage{bm}
\usepackage{color}
\usepackage{booktabs}
\usepackage{enumitem}
\usepackage{type1cm} 
\usepackage{mathrsfs}
\usepackage{comment}

\newcommand{\biblist}{\begin{list}{}
{\listparindent 0.0cm \leftmargin 0.50cm \itemindent -0.50 cm
\labelwidth 0 cm \labelsep 0.50 cm
\usecounter{list}}\clubpenalty4000\widowpenalty4000}
\newcommand{\ebiblist}{\end{list}}

\theoremstyle{plain}
\newtheorem{thm}{Theorem}
\newtheorem{cor}{Corollary}

\newcommand{\sep}{,\ }

\title{\bf Diagonals-parameter symmetry model and its property for square contingency tables with ordinal categories}
\author[1]{Kouji Tahata}
\author[2]{Kohei Matsuda}

\affil[1,2]{Department of Information Sciences, Tokyo University of Science}
\date{Last update: \today}

\begin{document}
\maketitle

\hrulefill
\begin{abstract}
Previously, the diagonals-parameter symmetry model based on $f$-divergence (denoted by DPS[$f$]) was reported to be equivalent to the diagonals-parameter symmetry model regardless of the function $f$, but the proof was omitted. Here, we derive the DPS[$f$] model and the proof of the relation between the two models. We can obtain various interpretations of the diagonals-parameter symmetry model from the result. Additionally, the necessary and sufficient conditions for symmetry and property between test statistics for goodness of fit are discussed.

\medskip

\noindent
{\bf Keywords}: Conditional symmetry\sep $f$-divergence\sep Global symmetry\sep Partial global symmetry

\medskip

\noindent
{\bf Mathematics Subject Classification}: 62H17 
\end{abstract}
\hrulefill 

\newpage
\section{Introduction}
\label{sec:intro}

A square contingency table with the same ordinal row and column categories may arise when a categorical variable is measured repeatedly.
In such a table, observations tend to concentrate on the main diagonal cells.
Our research aims to apply symmetry rather than the independence between row and column categories.
Many studies have treated symmetry issues, for example, \cite{B48}, \cite{KP97}, \cite{KA07}, \cite{TT11}, and \cite{T20}.

Let $X$ and $Y$ respectively denote the row and column variables for an $r \times r$ contingency table with ordinal categories.
Also, let $\pi_{ij}$ denote the probability that an observation falls in ($i,j$)th cell ($i=1,\dots,r;\ j=1,\dots,r$). 
\citet{G79b} proposed the diagonals-parameter symmetry (DPS) model, which is defined by
\begin{align}
  \label{eq:dps}
  \pi_{ij}=\left\{ \begin{array}{ll}
    d_k \psi_{ij} & (i<j), \\
    \psi_{ij} & (i\geq j), \\
  \end{array} \right.
\end{align}
where $\psi_{ij}=\psi_{ji}$ and $k=j-i$.
The parameter $d_{k}$ in the DPS model is simply the odds that an observation will fall in one of the cells $(i,j)$ where $j-i=k;\ i<j$, rather than in one of the cells $(j,i)$ where $j-i=k;\ i<j$ for $k=1,\dots,r-1$.
Additionally, for $j-i=k;\ i<j$, the ratio between $\pi_{ij}$ and $\pi_{ji}$ can be expressed the constant $d_{k}$.
That is, the ratio depends only on the distance from the main diagonal cells.

When equation (\ref{eq:dps}) with $d_1=\cdots=d_{r-1}=1$, the DPS model is reduced to the symmetry (S) model proposed by \citet{B48}.
When $d_k$ does not depend on $i$ or $j$ in equation (\ref{eq:dps}) with $d_1=\cdots=d_{r-1}$, the DPS model is reduced to the conditional symmetry (CS) model proposed by \citet{M78}.

\citet{KP97} described the DPS model based on the $f$-divergence (DPS[$f$]) model, which is defined as
\begin{align}
  \label{eq:dpsf}
  \pi_{ij}=\pi_{ij}^S F^{-1}(\Delta_k+\zeta_{ij}) \quad (i=1,\dots,r;\ j=1,\dots,r),
\end{align}
where $k=i-j$, $\pi^S_{ij}=(\pi_{ij}+\pi_{ji})/2$, $\zeta_{ij}=\zeta_{ji}$ and $\Delta_k+\Delta_{-k}=0$.
It should be noted that the function $f$ is a twice-differential and strictly convex, $F(t)=f^{\prime}(t)$, $f(1)=0$, $f(0)=\lim_{t\rightarrow 0}f(t)$, $0\cdot f(0/0)=0$, and $0\cdot f(a/0)=a\ {\rm{lim}}_{t\rightarrow\infty}[f(t)/t]$.
However, the derivation of this model is omitted in \citet{KP97}.
They also noted that (i) the DPS model is the closest model to symmetry in terms of the Kullback-Leibler (KL) distance and (ii) the DPS[$f$] model is equivalent to the DPS model.
In this study, we derive the DPS[$f$] model and the proof about the relation between the two models.
We can obtain various interpretations of the DPS model from the result.
Additionally, the necessary and sufficient condition for the S model and a property between test statistics for goodness of fit are discussed.

The rest of this paper is organized as follows.
Section \ref{sec:derivation} derives equation (\ref{eq:dpsf}) and interprets the model from an information theory viewpoint.
Additionally, the proof that the DPS[$f$] model is equivalent to the DPS model regardless of the function $f$ is given.
Section \ref{sec:separation} discusses the necessary and sufficient condition for the S model and highlights the relationships between the goodness-of-fit test statistics for the S model and the partitioned models.
Section \ref{sec:example} gives a numerical example.
Section \ref{sec:remarks} summarizes this paper.

\section{Properties of the DPS[$f$] model
\label{sec:derivation}}

\cite{KP97} noted that the DPS[$f$] model is the closest model to the S model in terms of the $f$-divergence under the conditions where ${\sum\sum}_{j-i=k} \pi_{ij}$ (and ${\sum\sum}_{i-j=k} \pi_{ij}$) for $k$=$1,\dots, r-1$ as well as the sums $\pi_{ij}+\pi_{ji}$ for $i=1,\dots,r;\ j=1,\dots r$, are given.
For example, \citet{IKK69}, \citet{KA07}, and \citet{T20} mentioned a similar property for the symmetry (or asymmetry) model.
This section derives the DPS[$f$] model and describes its properties.

We can obtain the following theorem, although the proof of Theorem \ref{thm:1} is given in Appendix.

\begin{thm}
  \label{thm:1}
  In the class of models with given ${\sum\sum}_{i-j=k} \; \pi_{ij}$, $k \neq 0$, and $\pi_{ij}+\pi_{ji}$ $(i=1,\dots,r;\ j=1,\dots,r)$, the model
\begin{align*}
  \pi_{ij}=\pi_{ij}^S F^{-1}(\Delta_k+\zeta_{ij}) \quad (i=1,\dots,r;\ j=1,\dots,r)
\end{align*}
with $k=i-j$, $\zeta_{ij}=\zeta_{ji}$ and $\Delta_k+\Delta_{-k}=0$, is the model closest to the complete symmetry model in terms of the $f$-divergence.
\end{thm}

The DPS$[f]$ model can be expressed as 
\begin{align}
  \label{eq:dpsf-2}
  F(2\pi_{ij}^c)=\left\{
  \begin{aligned}
  &\gamma_{ij}+a_k & (i<j), \\
  &\gamma_{ij} & (i\geq j), \\
  \end{aligned} \right.
\end{align}
where $k=j-i$, $\gamma_{ij}=\gamma_{ji}$ and $\pi_{ij}^c=\pi_{ij}/(\pi_{ij}+\pi_{ji})$.
Note that $\pi_{ij}^c$ is the conditional probability that an observation falls in the $(i,j)$th cell for a condition where the observation falls in the $(i,j)$th cell or the $(j,i)$th cell.
Namely, the DPS[$f$] model indicates that
\begin{align}
  \label{eq:dpsf-3}
  F(2\pi_{ij}^c) - F(2\pi_{ji}^c) = a_{k} \quad (i<j).
\end{align}
When $a_{1}=\cdots=a_{r-1}=0$, the DPS[$f$] model is reduced to the S model.

If $f(x)=x \log(x)$, $x>0$, then the $f$-divergence is reduced to the KL divergence.
When we set $f(x)=x\log(x)$, equation (\ref{eq:dpsf-2}) is reduced to
\begin{align*}
  \pi_{ij}=\left\{
    \begin{aligned}
    &\pi_{ij}^S\exp(\gamma_{ij}+a_k-1) & (i<j), \\
    &\pi_{ij}^S\exp(\gamma_{ij}-1) & (i\geq j), \\
  \end{aligned}
  \right.
\end{align*}
where $k=j-i$ and $\gamma_{ij}=\gamma_{ji}$.
We shall refer to this model as the DPS$_{\rm{KL}}$ model.
Under the DPS$_{\rm{KL}}$ model, the ratios of $\pi_{ij}$ and $\pi_{ji}$ for $i<j$ are expressed as
\begin{align}
  \label{eq:dps-2}
  \frac{\pi_{ij}}{\pi_{ji}}=d_{k}^{\rm{KL}} \quad (i<j),
\end{align}
where $d_{k}^{\rm{KL}}=\exp(a_k)$ and $k=j-i$.
Since equation (\ref{eq:dps-2}) indicates that the ratio of $\pi_{ij}$ and $\pi_{ji}$ depends on the distance of $k=j-i$, the DPS$_{\rm{KL}}$ model is equivalent to the DPS model proposed by \citet{G79b}.
Namely, the DPS model is the closest model to the S model in terms of the KL divergence under the conditions where ${\sum\sum}_{i-j=k} \; \pi_{ij}$, $k \neq 0$, and the sums $\pi_{ij}+\pi_{ji}$ for $i=1,\dots,r;\ j=1,\dots, r$ are given.
This is a special case of Theorem \ref{thm:1}.

If $f(x)=-\log(x)$, $x>0$, then the $f$-divergence is reduced to the reverse KL divergence.
Then, the DPS[$f$] model is reduced to
\begin{align*}
  \pi_{ij}=\left\{ 
    \begin{aligned}
    &\pi_{ij}^S \left( - \frac{1}{\gamma_{ij}+a_k} \right) & (i<j), \\
    &\pi_{ij}^S \left( - \frac{1}{\gamma_{ij}} \right) & (i\geq j), \\
\end{aligned} 
\right.
\end{align*}
where $k=j-i$ and $\gamma_{ij}=\gamma_{ji}$.
We shall refer to this model as the DPS$_{\rm{RKL}}$ model.
This model is the closest to the S model when the divergence is measured by the reverse KL divergence and can be expressed as
\begin{align*}
  \frac{1}{\pi_{ij}^{c}} - \frac{1}{\pi_{ji}^{c}} = d_{k}^{\rm{RKL}} \quad (i<j),
\end{align*}
where $d_{k}^{\rm{RKL}}=-2a_{k}$ and $k=j-i$.
This model indicates that the difference between inverse probabilities $1/\pi^{c}_{ij}$ and $1/\pi^{c}_{ji}$ depends on the distance of $k=j-i$.

If $f(x)=(1-x)^2$, then the $f$-divergence is reduced to the $\chi^{2}$-divergence (Pearsonian distance).
Then, the DPS[$f$] model is reduced to
\begin{align*}
  \pi_{ij}=\left\{ 
    \begin{aligned}
    &\pi_{ij}^S \left( \frac{\gamma_{ij}+a_k}{2}+1 \right) & (i<j), \\
    &\pi_{ij}^S \left( \frac{\gamma_{ij}}{2}+1 \right) & (i\geq j), \\
\end{aligned} 
\right.
\end{align*}
where $k=j-i$ and $\gamma_{ij}=\gamma_{ji}$.
We shall refer to this model as the DPS$_{\rm{P}}$ model.
This model is the closest to the S model when the divergence is measured by the $\chi^{2}$-divergence and can be expressed as
\begin{align*}
  \pi_{ij}^{c} - \pi_{ji}^{c} = d_{k}^{\rm{P}} \quad (i<j),
\end{align*}
where $d_{k}^{\rm{P}}=a_{k}/4$ and $k=j-i$.
This model indicates that the difference between $\pi^{c}_{ij}$ and $\pi^{c}_{ji}$ depends on the distance of $k=j-i$.

Moreover, if $f(x)=(\lambda(\lambda+1))^{-1}(x^{\lambda+1}-x)$, $x>0$, where $\lambda$ is a real-valued parameter, then the $f$-divergence is reduced to the power-divergence \citep{RC88}.
Then, the DPS$[f]$ model is reduced to
\begin{align*}
  \pi_{ij}=\left\{
    \begin{aligned}
    &\pi_{ij}^S \left( \lambda \left( \gamma_{ij}+a_k \right) +\frac{1}{\lambda+1} \right)^{\frac{1}{\lambda}} & (i<j),\\
    &\pi_{ij}^S \left( \lambda\gamma_{ij}+\frac{1}{\lambda+1} \right)^{\frac{1}{\lambda}} & (i\geq j), \\
  \end{aligned} 
  \right.
\end{align*}
where $k=j-i$ and $\gamma_{ij}=\gamma_{ji}$.
We shall refer to this model as the DPS$_{\rm{PD}(\lambda)}$ model.
This model is the closest to the S model when the divergence is measured by the power-divergence and can be expressed as
\begin{align*}
  (\pi_{ij}^{c})^{\lambda} - (\pi_{ji}^{c})^{\lambda} = d_{k}^{\rm{PD}(\lambda)} \quad (i<j),
\end{align*}
where $d_{k}^{\rm{PD}(\lambda)}=(\lambda a_{k})/2^{\lambda}$ and $k=j-i$.
This model indicates that the difference between the symmetric conditional probabilities to the power of $\lambda$ depends on the distance of $k=j-i$.
When we apply the DPS$_{\rm{PD}(\lambda)}$ model, we should set the value of $\lambda$.


\citet{KP97} reported that the DPS[$f$] model is equivalent to the DPS model regardless of $f$.
That is, the all models described above (i.e., DPS$_{\rm{KL}}$, DPS$_{\rm{RKL}}$, DPS$_{\rm{P}}$, and DPS$_{\rm{PD}(\lambda)}$) are equivalent to the DPS model surprisingly.
However, the proof was not given.
We prove the following theorem.
\begin{thm}
  \label{thm:2}
  The DPS[$f$] model is equivalent to the DPS model regardless of $f$.
\end{thm}

The poof is given in Appendix.
Theorem \ref{thm:2} states that the DPS model holds if and only if the DPS[$f$] model holds.
That is, if the DPS model fits the given dataset, then we obtain various interpretations for the data.

When $a_1=\cdots=a_{r-1}$, the DPS[$f$] model is reduced to the conditional symmetry model based on the $f$-divergence (CS[$f$]) model.
The CS[$f$] model is described previously \citet{KP97}.
Additionally, \citet{FT20} proposed the generalization of CS[$f$] model.
Similarly, when $d_{1}=\cdots=d_{r-1}$, the DPS model is reduced to the conditional symmetry (CS) model proposed by \citet{M78}.
The CS[$f$] model is equivalent to the CS model regardless of $f$ \citep{KP97}.
Hence, Theorem \ref{thm:2} leads to the following corollary. 

\begin{cor}
  \label{col:1}
  The CS[$f$] model is equivalent to the CS model regardless of $f$.
\end{cor}

\section{Equivalence conditions for symmetry
\label{sec:separation}}

Here, the equivalence conditions of the S model are discussed.
If the S model holds, then the DPS[$f$] model with $a_{1}=\cdots=a_{r-1}=0$ holds.
Conversely, if the DPS[$f$] model holds, then the S model does not hold generally.
Therefore, we are interested in considering an additional condition to obtain the S model when the DPS[$f$] model holds.
Other studies have discussed such conditions \citet{R77} and \citet{TNT16}.

We consider the distance global symmetry (DGS) model defined as
\begin{align}
  \label{eq:pgs}
  \delta^{U}_{k} = \delta^{L}_{k} \quad (k=1,\dots,r-1),
\end{align}
where $\delta^{U}_{k}={\sum\sum}_{j-i=k}\;\pi_{ij}$, $\delta^{L}_{k} = {\sum\sum}_{i-j=k}\;\pi_{ij}$.
For $k=1,\dots,r-1$, this model indicates that the sum of probabilities which are apart distance $k=j-i$ from main diagonal cells is equal to the sum of probabilities which are apart distance $k=i-j$ from main diagonal cells.
We obtain the following theorem. (The proof is given in Appendix.)

\begin{thm}
\label{thm:3}
The S model holds if and only if both the DPS[$f$] model and the DGS model hold.
\end{thm}

Next, we consider the global symmetry (GS) model, which is defined as
\begin{align*}
  \underset{i<j}{\sum\sum}\; \pi_{ij} = \underset{i<j}{\sum\sum}\; \pi_{ji}.
\end{align*}
It should be noted that the DGS model implies the GS model.
\citet{R77} noted that the S model holds if and only if both the CS model and the GS model hold.
\citet{FT20} proved that the S model holds if and only if the CS[$f$] model and the GS model hold.
These statements are the same as those from Corollary \ref{col:1}.


This section proves the separation of the test statistics for the S model into those for the DPS[$f$] model and the DGS model.
Let $n_{ij}$ denote the observed frequency in the $(i,j)$th cell in the $r \times r$ square contingency table.
Assume that a multinomial distribution applies to the $r\times r$ contingency table.
Let $m_{ij}$ and $\hat{m}_{ij}$ denote the expected frequency in the $(i,j)$th cell and the corresponding maximum likelihood estimate under a model, respectively.
Each model can be tested for the goodness of fit by, for example, the likelihood ratio chi-square statistic of model M, which is given as
\begin{align*}
  G^2(M)=2\sum_{i=1}^r\sum_{j=1}^rn_{ij}{\rm{log}}\left(\frac{n_{ij}}{{{\hat{m}}_{ij}}}\right),
\end{align*}
with the corresponding degree of freedom (df).

It is supposed that model M$_{3}$ holds if and only if both models M$_{1}$ and M$_{2}$ hold.
For these three models, \citet{A62} discussed the properties of the Wald test statistics, and \citet{DS63} described the properties of the likelihood ratio chi-square statistics.
Assume that the following equivalence holds:
\begin{align}
  \label{eq:orth}
  T(\mbox{M}_{3}) = T(\mbox{M}_{1}) + T(\mbox{M}_{2}),
\end{align}
where $T$ is the goodness of fit test statistic and the number of df for M$_{3}$ is equal to the sum of numbers of df for M$_{1}$ and M$_{2}$.
If both M$_{1}$ and M$_{2}$ are accepted with a high probability (at the $\alpha$ significance level), then M$_{3}$ is accepted.
However, when (\ref{eq:orth}) does not hold, an incompatible situation where both M$_{1}$ and M$_{2}$ are accepted with a high probability but M$_{3}$ is rejected may arise.
In fact, \citet{DS63} showed such an interesting example.

From Theorem \ref{thm:3}, the S model holds if and only if the DPS[$f$] model and the DGS model hold.
In addition, df for the DPS[$f$] model is $(r-1)(r-2)/2$ and that for DGS model is $(r-1)$.
Note that the number of df for the S model is equal to the sum of the numbers of df for the DPS[$f$] and the DGS models.
Thus, we consider partitioning test statistics.

Theorem \ref{thm:2} confirms that the DPS[$f$] model is equivalent to the DPS model.
Therefore, the maximum likelihood estimates (MLEs) under the DPS[$f$] model are given by
\begin{align}
  \label{eq:mle-dps}
  \left\{
    \begin{aligned}
      \hat{m}_{ij} &=\frac{n^{U}_k}{n^{U}_{k}+n^{L}_{k}} \left( n_{ij} + n_{ji} \right) & (i<j), \\
      \hat{m}_{ij} &=n_{ij} & (i=j),\\
      \hat{m}_{ij} &=\frac{n^{L}_{k}}{n^{U}_{k}+n^{L}_{k}} \left( n_{ij} + n_{ji} \right) & (i>j),
    \end{aligned}
  \right.
\end{align}
where $k=|j-i|$, $n^{U}_{k}=\sum\sum_{k=j-i} n_{ij}$, and $n^{L}_{k}=\sum\sum_{k=j-i} n_{ji}$ \citep{G79b}.

Next, we consider the MLEs under the DGS model using the Lagrange function.
Since the kernel of the log likelihood is $\sum_{i=1}^r\sum_{j=1}^rn_{ij} \log \pi_{ij}$, Lagrange function $L$ is written as 
\begin{align*}
  L=\sum_{i=1}^r\sum_{j=1}^rn_{ij}{\rm{log}} \; \pi_{ij} + \lambda \left( \sum_{i=1}^r\sum_{j=1}^r \pi_{ij}-1 \right) + \sum_{k=1}^{r-1} \lambda_k \left( \underset{k=j-i}{\sum\sum}\left(\pi_{ij}-\pi_{ji} \right) \right).
\end{align*}
Equating the derivation of $L$ to 0 with respect to $\pi_{ij}$, $\lambda$, and $\lambda_k$ gives
\begin{align}
  \label{eq:mle-pgs}
  \left\{
    \begin{aligned}
      \hat{m}_{ij} &= \frac{(n^{U}_{k}+n^{L}_{k})n_{ij}}{2n^{U}_{k}} & (i<j),\\
      \hat{m}_{ij} &= n_{ij} & (i=j),\\
      \hat{m}_{ij} &= \frac{(n^{U}_{k}+n^{L}_{k})n_{ij}}{2n^{L}_{k}} & (i>j),
    \end{aligned}
  \right.
\end{align}
where $k=|j-i|$.

We obtain the following equivalence from equations (\ref{eq:mle-dps}) and (\ref{eq:mle-pgs}).
\begin{align*}
  G^{2}(S) = G^{2}(DPS[f]) + G^{2}(DGS),
\end{align*}
because the MLEs under the S model are $\hat{m}_{ij}=(n_{ij}+n_{ji})/2$.
Therefore, the DPS[$f$] model and the DGS model are separable and exhibit independence.

Let $W(M)$ denote the Wald statistic for model M.
We obtain the following theorem and prove it in Appendix.

\begin{thm}
\label{thm:4}
$W(S)$ is equal to the sum of $W(DPS[f])$ and $W(DGS)$.
\end{thm}

\section{Numerical example\label{sec:example}}
Table \ref{tb:1}, which is taken from \citet{S06}, describes the amount of influence religious leaders and medical leaders should have in government funding for decisions on stem cell research when surveying 871 people.
The influence levels are divided into four categories: (1) Great influence, (2) Some influence, (3) A little influence, and (4) No influence.

\begin{table}[h]
\caption{\small How much influence should religious leaders and medical leaders have in government funding for decisions on stem cell research? \citep{S06}.\label{tb:1}}
\centering
\begin{tabular}{lllllr}\hline
Religious & \multicolumn{4}{c}{Medical Leaders} &\\\cline{2-5}
Leaders&Great(1)&Fair(2)&Little(3)&None(4)&Total\\\hline\hline
Great(1)&36&16&7&7&66\\
&$(36.00)^a$&(11.96)&(6.22)&(7.00)&\\
&$(36.00)^b$&(60.19)&(70.95)&(67.00)&\\
Fair(2)&74&96&22&4&196\\
&(78.04)&(96.00)&(26.05)&(4.78)&\\
&(42.67)&(96.00)&(82.76)&(40.55)&\\
Little(3)&119&174&48&4&345\\
&(119.78)&(169.95)&(48.00)&(3.99)&\\
&(62.59)&(100.34)&(48.00)&(15.05)&\\
None(4)&127&93&26&18&264\\
&(127.00)&(92.22)&(26.01)&(18.00)&\\
&(67.00)&(48.91)&(14.99)&(18.00)&\\ \hline
Total&356&379&103&33&871\\\hline
\multicolumn{6}{l}{${}^{a}$MLEs under the DPS model}\\
\multicolumn{6}{l}{${}^{b}$MLEs under the DGS model}\\
\end{tabular}
\end{table}

Table \ref{tb:2} gives the values of the likelihood ratio chi-square statistics $G^2$ and $p$ value for the models applied to these data. 
Table \ref{tb:2} indicates that the sum of the test statistics DPS (i.e., DPS[$f$]) model and DGS model is equal to that of the S model.
The S model fits the data very poorly.
We can infer that the marginal distribution for religious leaders is not equal to that for medical leaders.
On the other hand, the DPS model fits the data very well.
Additionally, the DGS model fits the data poorly.
From Theorem \ref{thm:3}, the reason of the poor fit of S model is caused by the poor fit of DGS model rather than the DPS model.

\begin{table}[h]
\caption{\small Likelihood ratio chi-square values $G^2$ for the models applied to Table \ref{tb:1}. \label{tb:2}}
\centering
\begin{tabular}{lrrr}\hline
Models &df & $G^2$ & $p$-value\\\hline\hline
S&6&545.15&$<0.0001$\\
DPS&3& 2.45&0.4847\\
DGS&3&542.70&$<0.0001$\\\hline
\end{tabular}
\end{table}

The values of MLEs of $(d_1,d_2,d_3)$ in equation (\ref{eq:dps}) are $(0.15, 0.05, 0.06)$.
It should be noted that $(d_1,d_2,d_3)$ is equal to $(d_{1}^{\rm{KL}},d_{2}^{\rm{KL}},d_{3}^{\rm{KL}})$ in the DPS$_{\rm{KL}}$ model.
Let $(i,j)$ denote the pair that the amount of influence religious leaders is $i$th level and that of medical leaders is $j$th level.
When $k=j-i$ ($k=1,2,3$), a pair $(i,j)$ is $\hat{d}_{k}$ times as likely as a pair $(j,i)$ on condition that a pair is $(i,j)$ or $(j,i)$.
From $\hat{d}_{k} < 1$ ($k=1,2,3$), the probability distribution for religious leaders is {\it stochastically higher} than the probability distribution of medical readers.
That is, the medical leaders rather than the religious leaders should have influence in government funding for decisions on stem cell research.

Moreover, from Theorem \ref{thm:2}, we can obtain various interpretations.
Since the DPS model holds, the DPS$_{\rm{RKL}}$, DPS$_{\rm{P}}$, and DPS$_{\rm{PD}(\lambda)}$ models also hold.
For example, we obtain
\begin{align*}
(\hat{d}_{1}^{\rm{RKL}},\hat{d}_{2}^{\rm{RKL}},\hat{d}_{3}^{\rm{RKL}}) &= (6.37, 19.22, 18.09),\\
(\hat{d}_{1}^{\rm{P}},\hat{d}_{2}^{\rm{P}},\hat{d}_{3}^{\rm{P}}) &= (-0.73, -0.90, -0.90),
\end{align*}
and for $\lambda=3$,
\begin{align*}
(\hat{d}_{1}^{\rm{PD}(3)},\hat{d}_{2}^{\rm{PD}(3)},\hat{d}_{3}^{\rm{PD}(3)}) = (-0.65, -0.86, -0.85).
\end{align*}
When $k=j-i$ ($k=1,2,3$), we can infer that (i) the difference between the reciprocal of conditional probability that a pair is $(i,j)$ and the reciprocal of conditional probability that a pair is $(j,i)$ is $\hat{d}_{k}^{\rm{RKL}}$ on condition that the pair is $(i,j)$ or $(j,i)$ from the DPS$_{\rm{RKL}}$ model, (ii) the difference between the conditional probability that a pair is $(i,j)$ and the conditional probability that a pair is $(j,i)$ is $\hat{d}_{k}^{\rm{P}}$ under the same condition from the DPS$_{\rm{P}}$ model, and (iii) the difference between the conditional probability that a pair is $(i,j)$ to the third power and the conditional probability that a pair is $(j,i)$ to the third power is $\hat{d}_{k}^{\rm{PD}(3)}$ under the same condition from the DPS$_{\rm{PD}(3)}$ model.

\section{Concluding remarks\label{sec:remarks}}
This paper provides the proof that the DPS[$f$] model is equivalent to the DPS model proposed by \citet{G79b}.
This result provides the various interpretations of the DPS model.
Additionally, the separation of the test statistic for the S model is considered.
The DPS[$f$] model and the DGS model is separable and exhibit independence.
\citet{KP97}, \citet{KA07}, \citet{T20} and \citet{FT20} considered models based on the $f$-divergence for the analysis of square contingency tables with ordinal categories.
In the future, whether the model based on the $f$-divergence is equivalent to the conventional model should be studied.

\section*{Acknowledgments}
This work was supported by JSPS KAKENHI (Grant Number 20K03756).

\newpage        
\section*{Appendix}
\label{sec.Ap}
This section provides the proofs of theorems.

\subsection*{Proof of Theorem \ref{thm:1}}

We note that ($\pi_{ij}^S$) satisfies the symmetry structure for cell probabilities and is given under the conditions where
\begin{align}
\label{eq:r1}
  \pi_{ij}+\pi_{ji}=t_{ij}=t_{ji} \quad (i=1,\dots,r;j=1,\dots,r)
\end{align}
are given.
Let $I^C(\pi :\pi^S)$ denote the $f$-divergence between ($\pi_{ij}$) and ($\pi_{ij}^S$).
That is
\begin{align}
\label{eq:f-d}
  I^C(\pi :\pi^S)=\sum_{i=1}^r\sum_{j=1}^r\pi_{ij}^Sf\left(\frac{\pi_{ij}}{\pi_{ij}^S} \right),
\end{align}
where $f$ satisfies the conditions described in Section \ref{sec:intro}.
Now minimize (\ref{eq:f-d}) under the conditions where the restraints (\ref{eq:r1}) in addition to
\begin{align}
\label{eq:r2}
  \delta^U_{-k}=\underset{i-j=-k}{\sum\sum} \; \pi_{ij} \quad {\rm and} \quad \delta^L_k=\underset{i-j=k}{\sum\sum} \; \pi_{ij} \quad (k=1,\dots,r-1)
\end{align}
are given.
Note that (i) $I^C(\pi:\pi^S)$ is strictly convex and (ii) the restrictions are linear equations. The Lagrange function is written as
\begin{multline*}
  L=I^C(\pi:\pi^S)+\sum_{i=1}^r\sum_{j=1}^r\lambda_{ij} \left( \pi_{ij}+\pi_{ji}-t_{ij} \right)\\
  +\sum_{k=1}^{r-1} \left( \bar{\Delta}_{-k} \left( \underset{i-j=-k}{\sum\sum}\;\pi_{ij}-\delta^U_{-k} \right) + \bar{\Delta}_{k}\left( \underset{i-j=k}{\sum\sum}\;\pi_{ij}-\delta^L_k \right) \right) .
\end{multline*}
Equating derivation $L$ to 0 with respect to $\pi_{ij}$ gives
\begin{align}
  \label{eq:r3}
  \left\{
  \begin{aligned}
  f^{\prime}\left(\frac{\pi_{ij}}{\pi_{ij}^S}\right) &+ \bar{\Delta}_{-k} + \lambda_{ij}+\lambda_{ji}=0 & (i<j),\\
  f^{\prime}\left(\frac{\pi_{ij}}{\pi_{ij}^S}\right) &+ \lambda_{ij}+\lambda_{ji}=0 & (i=j),\\
  f^{\prime}\left(\frac{\pi_{ij}}{\pi_{ij}^S}\right) &+ \bar{\Delta}_{k} + \lambda_{ij}+\lambda_{ji}=0 & (i>j).\\
  \end{aligned}
  \right.
\end{align}
Let $f^{\prime}$ denote $F$, and let $\pi_{ij}^{\ast}$ denote the solution satisfying (\ref{eq:r1}), (\ref{eq:r2}), and (\ref{eq:r3}).
Since $f$ is a strictly convex function, it follows that $F^{\prime}(x)=f^{{\prime}{\prime}}(x)>0$ for all $x$.
Hence, $F$ is strictly monotone and ensures that $F^{-1}$ exists.
Let $-(\lambda_{ij}+\lambda_{ji})$ and $-\bar{\Delta}_l$ denote $\zeta_{ij}$ and $\Delta_l$, respectively.
From equation (\ref{eq:r3}), we obtain
\begin{align*}
  \left\{
  \begin{aligned}
  \pi^{\ast}_{ij} &= \pi_{ij}^S F^{-1} \left(\Delta_{-k} + \zeta_{ij} \right) & (i<j),\\
  \pi^{\ast}_{ij} &= \pi_{ij}^S F^{-1} \left(\zeta_{ij}\right) & (i=j),\\
  \pi^{\ast}_{ij} &= \pi_{ij}^S F^{-1} \left(\Delta_{k} + \zeta_{ij} \right) & (i>j),\\
  \end{aligned}
  \right.
\end{align*}
where $\zeta_{ij}=\zeta_{ji}$ and $\Delta_{k}+\Delta_{-k}=0$.
The minimum value of $I^C(\pi:\pi^S)$ is attained for $\pi_{ij}^{\ast}$ where $\zeta_{ij}$ and $\Delta_l$ are determined so that $\pi_{ij}^{\ast}$ satisfies restraints (\ref{eq:r1}) and (\ref{eq:r2}).
Therefore, the DPS[$f$] model is the closest model to the S model in terms of the $f$-divergence under these conditions.

\subsection*{Proof of Theorem \ref{thm:2}}
  Let function $G$ be defined as
  \begin{align*}
    G(x) = F\left( \frac{2x}{1+x} \right) - F\left( \frac{2}{1+x} \right) \quad (x > 0),
  \end{align*}
  where $F=f^{\prime}$.
  Then, the derivative of $G$ is 
  \begin{align*}
    G^{\prime}(x) = \frac{2}{(1+x)^2}\left(F^{\prime}\left(\frac{2x}{1+x}\right)+F^{\prime}\left(\frac{2}{1+x}\right)\right).
  \end{align*}
  Since the function $f$ is twice-differential and strictly convex that $G^{\prime}(x)>0$ for $x>0$.
  Hence, $G$ is a strictly increasing function, and $G^{-1}$ exists.
  
  If the DPS model holds, $\pi_{ij}/\pi_{ji} = d_{k}$ holds for $i<j$ from equation (\ref{eq:dps}), where $k=j-i$.
  Then we can see that for $i<j$,
  \begin{align*}
    G(d_{k}) &=  F\left( \frac{2d_{k}}{1+d_{k}} \right) - F\left( \frac{2}{1+d_{k}} \right), \\
    &= F\left( 2 \pi_{ij}^{c} \right) - F\left( 2 \pi_{ji}^{c} \right).
  \end{align*}
  This is equivalent to equation (\ref{eq:dpsf-3}).
  Namely, the DPS[$f$] model holds.

  On the other hand, if the DPS[$f$] model holds, equation (\ref{eq:dpsf-3}) holds.
  We can see that for $i<j$,
  \begin{align*}
    G\left( \frac{\pi_{ij}}{\pi_{ji}} \right) = a_{k}.
  \end{align*}
  Since $G^{-1}$ exists, we obtain
  \begin{align*}
    \frac{\pi_{ij}}{\pi_{ji}} = G^{-1} \left( a_{k} \right).
  \end{align*}
  Namely, the DPS model holds.
  The proof is complete.

\subsection*{Proof of Theorem \ref{thm:3}}
It is obvious that if the S model holds, the DPS[$f$] model and the DGS model simultaneously hold.
Assuming that both the DPS[$f$] model and the DGS model hold, we show that the S model holds.
From Theorem \ref{thm:2}, the DPS[$f$] model is equivalent to $\pi_{ij}/\pi_{ji}=d_{k}$ for $i<j$ with $k=j-i$.
Since the DGS model holds, we obtain
\begin{align*}
  \underset{j-i=k}{\sum\sum}\; \left(d_{k} - 1\right)\pi_{ji} = 0 \quad (k=1,\dots,r-1).
\end{align*} 
Since $\pi_{ji} >0$, we get $d_k=1$ ($k=1,\dots,r-1$).
Namely, the S model holds.

\subsection*{Proof of Theorem \ref{thm:4}}
Theorem \ref{thm:2} shows that the DPS[$f$] model is equivalent to the DPS model.
Let
\begin{align*}
\bm{\pi} &=(\pi_{11},\dots,\pi_{1r},\pi_{21},\dots,\pi_{2r},\dots,\pi_{r1},\dots,\pi_{rr})^t,\\
\bm{\beta} &=(\rho_1,\dots,\rho_{r-1},\bm{\varepsilon})^t,
\end{align*}
where $\bm{\varepsilon}=(\varepsilon_{11},\dots,\varepsilon_{1r},\varepsilon_{22},\dots,\varepsilon_{2r},\dots,\varepsilon_{rr})$.
Then, from equation (\ref{eq:dps}), the DPS model is expressed as
\begin{align}
\label{q17}
\log\bm{\pi}=\bm{X\beta}=(\bm{x}_1,\dots,\bm{x}_{r-1},\bm{x}_{11},\dots,\bm{x}_{1r},\bm{x}_{22},\dots,\bm{x}_{2r},\dots,\bm{x}_{rr})\bm{\beta},
\end{align}
where $\bm{x}_l=(\bm{w}_{l+1},\dots,\bm{w}_{r},0,\dots,0)^t$ is a $r^2 \times 1$ vector $(l=1,\dots,r-1)$.
Here, $\bm{w}_h$ ($1 \times r$ vector) is 1 for the $h$th element and 0 otherwise.
For example, when $r=4$, 
\begin{align*}
\bm{x}_1=(\bm{w}_{2},\bm{w}_{3},\bm{w}_{4},0,\dots,0)^t=(0,1,0,0,0,0,1,0,0,0,0,1,0,0,0,0)^t.
\end{align*} 
Additionally, $\bm{x}_{ij}$ ($i \leq j$) is the $r^{2} \times 1$ vector shouldering $\varepsilon_{ij}$.
Note that the $r^2\times K$ matrix $\bm{X}$ is a full column rank where $K=(r-1)+r(r+1)/2$.

We denote the linear space spanned by the column of matrix $\bm{X}$ by $S(\bm{X})$ with dimension $K$.
$S(\bm{X})$ is a subspace of $\mathbb{R}^{r^2}$.
Let $U$ be an $r^2\times d_1$ full column rank matrix such that the linear space $S(\bm{U})$ spanned by the column of $\bm{U}$ is the orthogonal complement of the space $S(\bm{X})$.
Note that $d_1=r^{2}-((r-1)+r(r+1)/2)=(r-1)(r-2)/2$.
Since $\bm{U}^t\bm{X}=\bm{O}_{d_1,K}$ where $\bm{O}_{d_1,K}$ is the $d_1 \times K$ zero matrix, the DPS model can be expressed as
$\bm{h}_1(\bm{\pi})=\bm{U}^t\log\bm{\pi}=\bm{0}_{d_1}$, where $\bm{0}_s$ is the $s \times 1$ zero vector.

Additionally, the DGS model can be expressed as $\bm{h}_2(\bm{\pi})=\bm{M\pi}=\bm{0}_{d_2}$ where
\begin{align*}
\bm{M}=(\bm{g}_1,\dots,\bm{g}_{r-1})^t,
\end{align*}
and $d_2=r-1$.
Here, $\bm{g}_l = 2\bm{x}_l-\sum\sum_{j-i=l}\bm{x}_{ij}$.
Note that $\bm{M}^t$ belongs to the space $S(\bm{X})$.
That is, $S(\bm{M^t}) \subset S(\bm{X})$.

Let $\bm{p}$ denote $\bm{\pi}$ with $\pi_{ij}$ replaced by $p_{ij}$, where $p_{ij}=n_{ij}/n$ with $n=\sum\sum n_{ij}$.
From Theorem \ref{thm:3}, the S model is equivalent to $\bm{h}_3(\bm{\pi})=\bm{0}_{d_3}$, where $\bm{h}_3=(\bm{h}_1^t,\bm{h}_2^t)^t$ and $d_3=d_1+d_2=r(r-1)/2$.
In an analogous manner to \citet{T20}, we obtain that $\sqrt{n}(\bm{h}_3(\bm{p})-\bm{h}_3(\bm{\pi}))$ has an asymptotically normal distribution with mean $\bm{0}_{d_3}$ and covariance matrix
\begin{align*}
\bm{H}_3(\bm{\pi}) \bm{\Sigma}(\bm{\pi}) \bm{H}_3^t(\bm{\pi})= \left[
\begin{array}{rr}
\bm{H}_1(\bm{\pi}) \bm{\Sigma} (\bm{\pi} ) \bm{H}_1^t(\bm{\pi}) & \bm{O}_{d_1,d_2} \\
\bm{O}_{d_2,d_1} & \bm{H}_2(\bm{\pi}) \bm{\Sigma} (\bm{\pi} ) \bm{H}_2^t(\bm{\pi}) \\
\end{array}
\right],
\end{align*}
where $\bm{H}_s(\bm{\pi})=\partial \bm{h}_s(\bm{\pi})/\bm{\pi}^t$ and $\bm{\Sigma} (\bm{\pi} )=diag(\bm{\pi})-\bm{\pi}\bm{\pi}^{t}$.
Here, $diag(\bm{\pi})$ denotes a diagonal matrix with the $i$th component of $\bm{\pi}$ as the $i$th diagonal component.
Therefore, $W_3=W_1+W_2$ holds, where
\begin{align*}
W_s=n\bm{h}^t_s(\bm{p}) (\bm{H}_s(\bm{p}) \bm{\Sigma} (\bm{p} ) \bm{H}_s^t(\bm{p}))^{-1} \bm{h}_s(\bm{p}).
\end{align*}
The Wald statistic for the DPS[$f$] model (i.e., $W(DPS[f])$) is $W_1$, that for the DGS model (i.e., $W(DGS)$) is $W_2$, and that for the S model (i.e., $W(S)$) is $W_3$.
The proof is complete.

    


\newpage
\bibliographystyle{duc1} 
\bibliography{tahata_refs}

\end{document}